\documentstyle[twoside, epsf]{article}

\input ibvs2.sty

\begin{document}
\IBVShead{xxxx}{xx xxxxxx xxxx}

\IBVStitletl{New Light Curves and Ephemeris for the}{Close Eclipsing Binary V963 Per}
 
\begin{center}
\IBVSauth{ODELL, ANDREW P.$^1$; WILS, PATRICK$^2$; DIRKS, CLARISSA$^3$; GUVENEN, BLYTHE$^4$; O'MALLEY, C. JO$^4$; 
VILLARREAL, ANTONIO S.$^4$; WEINZETTLE, RITA M.$^4$}
\end{center}

\IBVSinst{Dept of Physics and Astronomy, NAU Box 6010, Flagstaff AZ 86011 USA; e-mail: WCorvi @ yahoo.com}
\IBVSinst{Vereniging Voor Sterrenkunde, Belgium}
\IBVSinst{Evergreen State College, Olympia WA USA}
\IBVSinst{Steward Observatory, University of Arizona, Tucson AZ 85721 USA}

\SIMBADobj{GSC 03355 00394}
\SIMBADobj{W Crv}
\GCVSobj{V963 Per}

\IBVSkey{Binaries:eclipsing}

\IBVSabs{We present new light curves for V963 Per and derive a new ephemeris. An abrupt change }
\IBVSabs{in the depth of secondary minimum of about 4$\%$ occurred during the observations.}

\begintext

V963 Per (GSC 3355 0394, $\alpha$(2000)=04\hr 45\mm 35\fsec6, $\delta$(2000)=+52\deg 22\arcm 35\farcs4) is a close eclipsing binary identified by Nicholson and Varley (Anonymous 2006) and was recently discussed extensively by Samec et al. (2010, hereafter referred to as RGS), who obtained BVRI light curves on two nights.  This star is potentially interesting astrophysically because of its short period but large difference between eclipse depths, possibly similar to W Corvi (see, eg Odell \& Cushing 2004).  RGS fit their light curves with a model having three spots on the secondary and one on the primary star. RGS also found a large mass ratio (q = $\sim$ 0.87), which gives the possibility of detecting the secondary star in spectral line profiles, thus revealing the spot characteristics.   There are some inconsistencies in RGS, but, our new photometry confirms the basic shape of the light variation that RGS found, and the system remains an interesting one.

We have obtained new light curves in 2010-2011 at the 1.55m Kuiper telescope of the Steward Observatory utilizing the Mont4KCCD camera binned 3x3 (described in detail by Randall et al. 2007) utilizing Bessell-B, -V, -R and Arizona-I filters.  Data reduction was done in the following way: The two amplifier readouts were corrected for overscan and crosstalk on the fly using an IRAF script.  Then the images were Zero and Flat Field corrected using IRAF*, and magnitudes were extracted with qphot task with 4\farcs2 radius apertures.  

The Zero and Flat Field frames were the average of 200 images each, pointed at a dome screen such that $\sim$ 10K counts was reached for each flat.  In order to minimize any flat-fielding problems, an autoguider kept all stars on the same pixels for the entire run, and focus was monitored carefully.  To check for possible color terms in the extinction, the third comparison star (with very similar color to the eclipser) was treated as `variable' and its (V-C) magnitude was plotted as a function of airmass.  No significant effect was apparent (i.e. $<$ 0.002 mag).

The comparison and check stars were treated in an unusual way, in that a `combined comparison' was formed from the average of five bright stars surrounding the variable (see Table 1).  This has two advantages: it improves the S/N of the comparison, and it better accounts for any variations in transparency over the field, as opposed to using only one star.  It also makes any anomalous reading immediately apparent.  In order to compare our data to that of RGS, we then subtracted from each magnitude difference (Variable - Comparison) the filter-specific difference between RGS's comparison star and our average of five stars, averaged over each night, so our results are on the same scale as theirs.

\begin{table}[!ht]
\small
\centerline{{\bf Table 1.} Comparison Stars}
\begin{center}
\begin{tabular}{crcl}
\hline
   GSC    &  Offset & Offset  & comment \\
 number   &  RA     & dec     &         \\
\hline
3355 0474 &  -3\fsec1   & -3\arcm07\arcs &         \\	
3355 0096 & -20\fsec9   & -2\arcm06\arcs &         \\
3355 0362 &  15\fsec3   & +1\arcm01\arcs & same color as variable\\
3355 0336 & -27\fsec1   & +3\arcm04\arcs &         \\
3355 0596 &   7\fsec4   & +4\arcm43\arcs & RGS comparison star \\
\hline
\end{tabular}
\end{center}
\end{table}



We plan to make our observations available as Table 2: HJD, orbital phase (based on Eq. 1) and delta magnitude: (Variable minus Comparison star). The table will be available through the IBVS website as {\tt xxxx-t2.txt}.

\IBVSedata{xxxx-t2.txt}
\IBVSdataKey{xxxx-t2.txt}{V963 Per}{photometry}

Fig. 1 shows the complete dataset for our light curves.  The number given in the key for each filter represents the magnitude difference required to correct to RGS's comparison star.  It can be seen that the light curves do not reproduce perfectly - there is a small night-to-night scatter around phase 0.20, and a larger effect at phases between 0.40 and 0.90.  The former could conceivably be due to a slightly incorrect period being used for the phasing (see the discussion of the ephemeris later in this paper); the latter must be caused by intrinsic variation of the star, as would general differences in level between RGS and our new data.  The O'Connell effect is rather large; phase 0.75 is about 0.10 mag fainter than phase 0.25.

Five new times of primary minimum and two of secondary were obtained, which alter the ephemeris.  The variations in the light curve near phase 0.5 must be real, as this is during secondary eclipse, which is obviously total.  Fig. 2 shows the relevant part of the cycle in the B filter in detail; data from before January 1, 2011 is plotted as diamonds and after that date as triangles.  The depth of secondary eclipse became fainter by about 0.04 mags somewhere around that date.  The change extends from about phase 0.40 to about 0.80, which includes the time when the secondary star is completely eclipsed.

RGS present their data in Table 1 and Fig. 2 of that paper; although the figures seem to be correct, the magnitudes listed in the table for the R and I filters appear to be fluxes, since the value at maximum (phase 0.25) is exactly 1.0, and the light curves would be inverted if the numbers are correct.  We also note that the HJD given for all filters is likely erroneous (see below).  However, there seems to be a difference of about 0.1 mag at all phases between RGS's data and ours (in the sense they found the star to be fainter).  The exact cause of this isn't clear, since our images didn't include RGS's check star.

We used the ephemeris of RGS to predict that a primary eclipse would be seen on our first night of observation, but in fact, secondary eclipse was observed, indicating a problem there.  A new, preliminary ephemeris was determined from our five primary timings given in Table 3 to be HJD = 2455563.6833(2) + 0.462087(3) (used to phase the data in our figures 1 and 2).  We determined our times of minimum by folding our data to see where the ascending light best fell on the descending branch; our uncertainties come from this measure for the four different filters (standard deviation).  The short time (200 cycles) over which the data were obtained makes this period rather uncertain, but definitely different from RGS's 0.46216 days.

\IBVSfig{10.5cm}{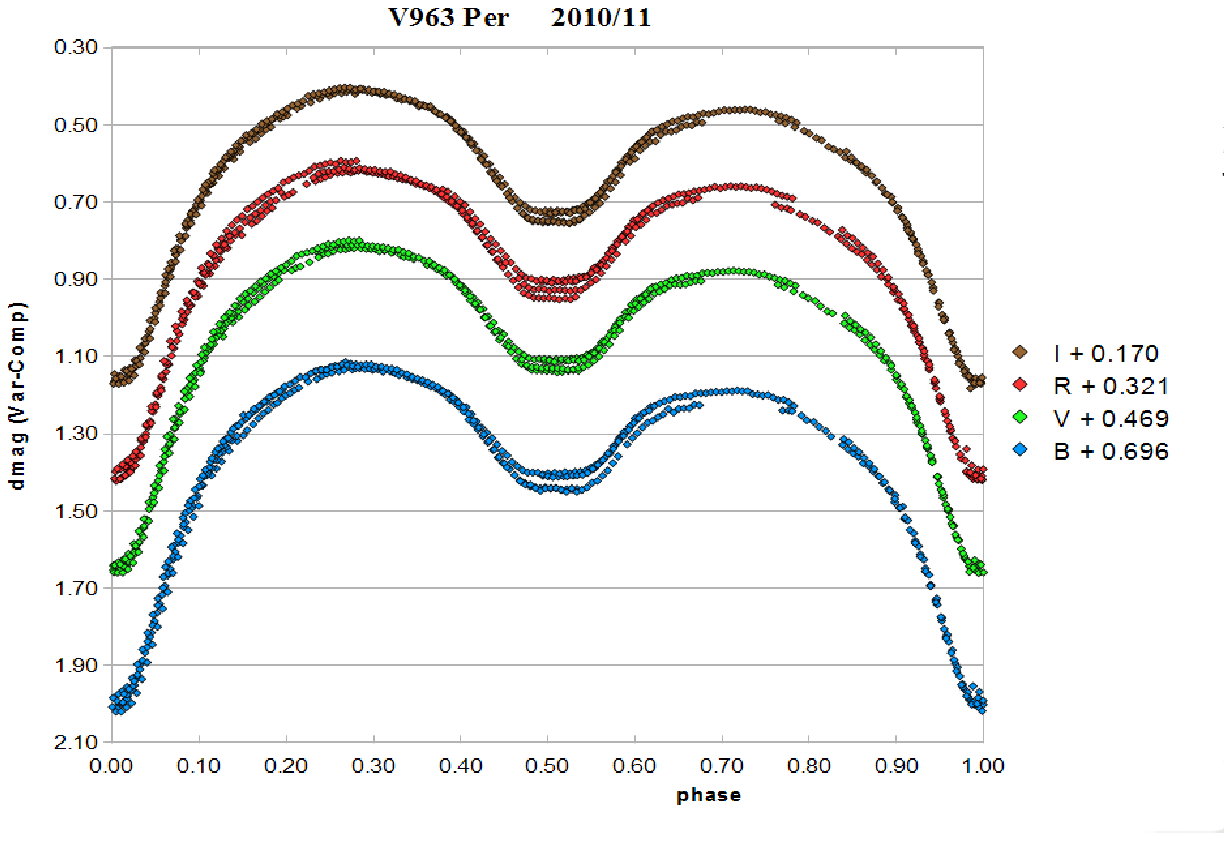}{Light Curves for V963 Per in 2010-11.}

\IBVSfig{10.cm}{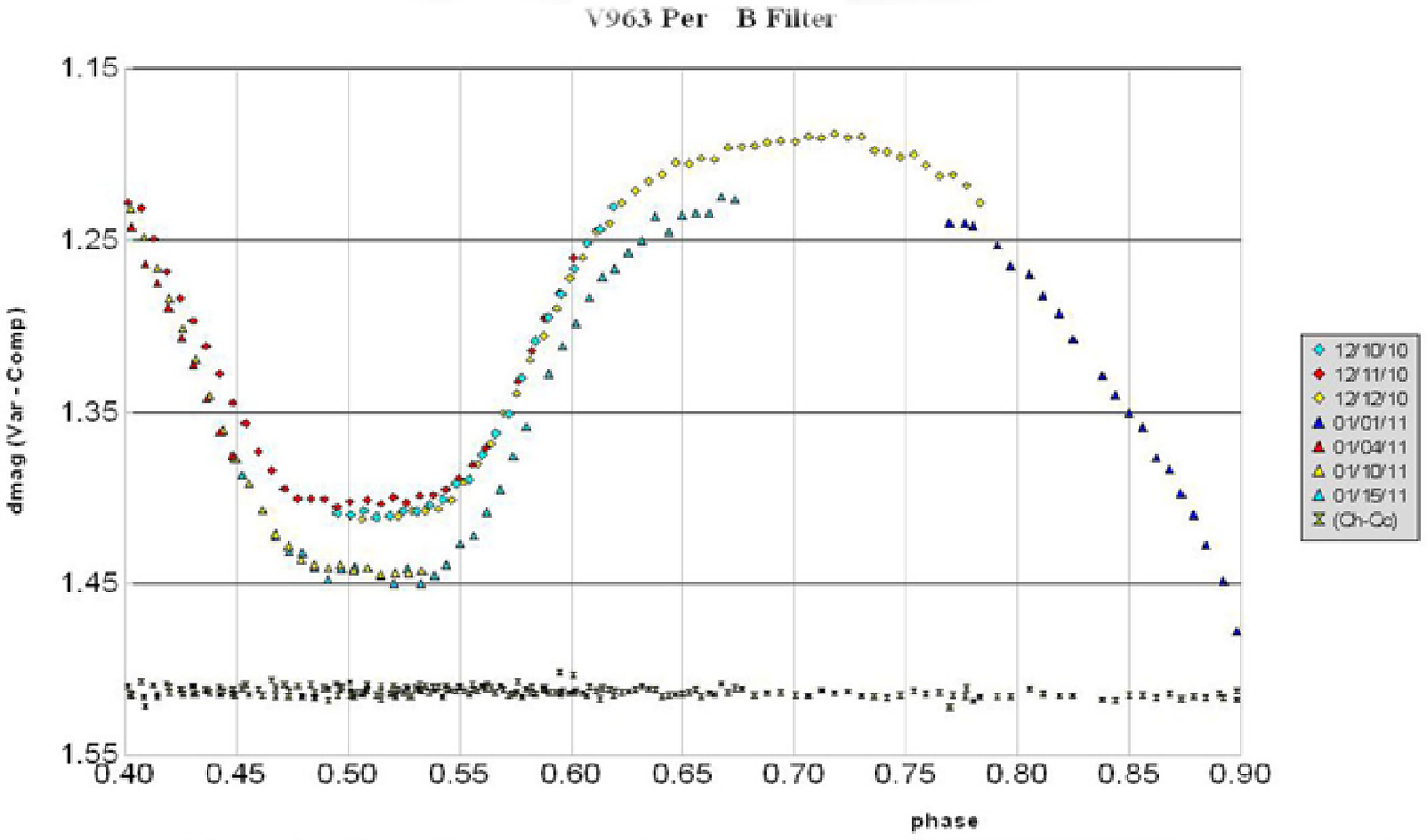}{Variation of the Light Curve for V963 Per in B Filter.  The points at dmag 1.51 are (Five-star-comp - Comp3) which would show any variability in the comparison stars.}

\vspace*{-0.85cm}

\begin{table}[!ht]
\small
\centerline{{\bf Table 3.} $O-C$ Linear Residuals, Eq. 1}
\begin{center}
\begin{tabular}{cccrcl}
\hline
	  No  &      Epoch       & Uncertainty &    Cycle    &   (Obs-Calc)  &  Comment\\
        &       HJD        &    days     &             &    minutes    &         \\
\hline
    1   &    2455542.6626  &   0.0005    &    -45.5    &      5.52     &   Secondary - not used\\
    2   &    2455563.6835  &   0.0001    &      0.0    &      0.23     &   Primary\\
    3   &    2455564.6077  &   0.0001    &      2.0    &      0.33     &   Primary\\
    4   &    2455576.6212  &   0.0002    &     28.0    &     $-$0.47   &   Primary\\
    5   &    2455577.7818  &   0.0006    &     30.5    &      7.32     &   Secondary - not used\\
    6   &    2455601.5748  &   0.0013    &     82.0    &      1.46     &   Primary\\
    7   &    2455618.6717  &   0.0002    &    119.0    &      1.55     &   Primary\\
\hline
\end{tabular}
\end{center}
\vspace*{-0.4cm}
\end{table}

\vspace*{1.0cm}

When our new period was used to compute times of minimum to compare to RGS's timings, a problem developed.  RGS tells of two nights' data acquisition, but gives timings on four nights, two in November 2007, one in December 2008, and one in January 2009 (according to the Julian Dates).  However, our current period yields residuals for the last four of RGS's timings of over two hours.  The times of observation listed in RGS, Table 1, do not correspond to the nights they claimed to have observed this star.  Thus we must hold off on using any of those timings in attempting to derive a better period or to test for period variation.

Instead, we use the data from the Northern Sky Variability Survey (NSVS, see Wozniak et al. 2004) taken in 2000 and 2001 and from the SuperWASP (see Butters et al. 2010) dataset taken in 2007 to improve the ephemeris.  Each of these datasets contain only a few measurements on any one night, so the following strategy was adopted.  We used OpenOffice Calc to calculate phases for all unflagged measurements in the two datasets based on a chosen linear ephemeris.  We also calculated the predicted timings for all minima measured here and by RGS.  The initial epoch was fixed on the value derived for our five primary minima.  The period was varied so as to maintain small residuals for the most recent data (sensitive to the initial epoch but not the period), and to get a qualitatively good light curve for the NSVS and SuperWASP data sets (sensitive to the period, but not the initial epoch).  This technique completely avoids any cycle count ambiguity, but precludes putting uncertainties on the results.

\begin{equation}
  {\rm HJD\ T_{min}\ I} = 2455563.6833 + 0.462078{\rm d} \times {\rm E}.
\end{equation}
\vspace*{-0.2cm}

The final ephemeris is given by Eq. 1.  The residuals for our timings are given in Table 3, and the plots of the two earlier datasets are given in Fig. 3.  Note that primary minimum in the NSVS data seems to be near phase 0.96, and in the SuperWASP data at phase 0.03.  This constrains the period rather well; improving either of these datasets makes the other less compelling.  The discrepancy of each at about 0.04 cycles could be due to inaccuracies in those datasets (there is large intrinsic scatter, for example), or could be indicating a quadratic ephemeris.  Also note in Table 3 that the secondary eclipses have residuals of about six minutes, which indicates that they can not be used (and weren't used) to derive the ephemeris.  This is also obvious from Fig, 2, where it can be seen that the change in secondary minimum is more prominent after phase 0.50 than before; time of secondary minimum is clearly affected by the variation of the light curve.

\IBVSfig{7cm}{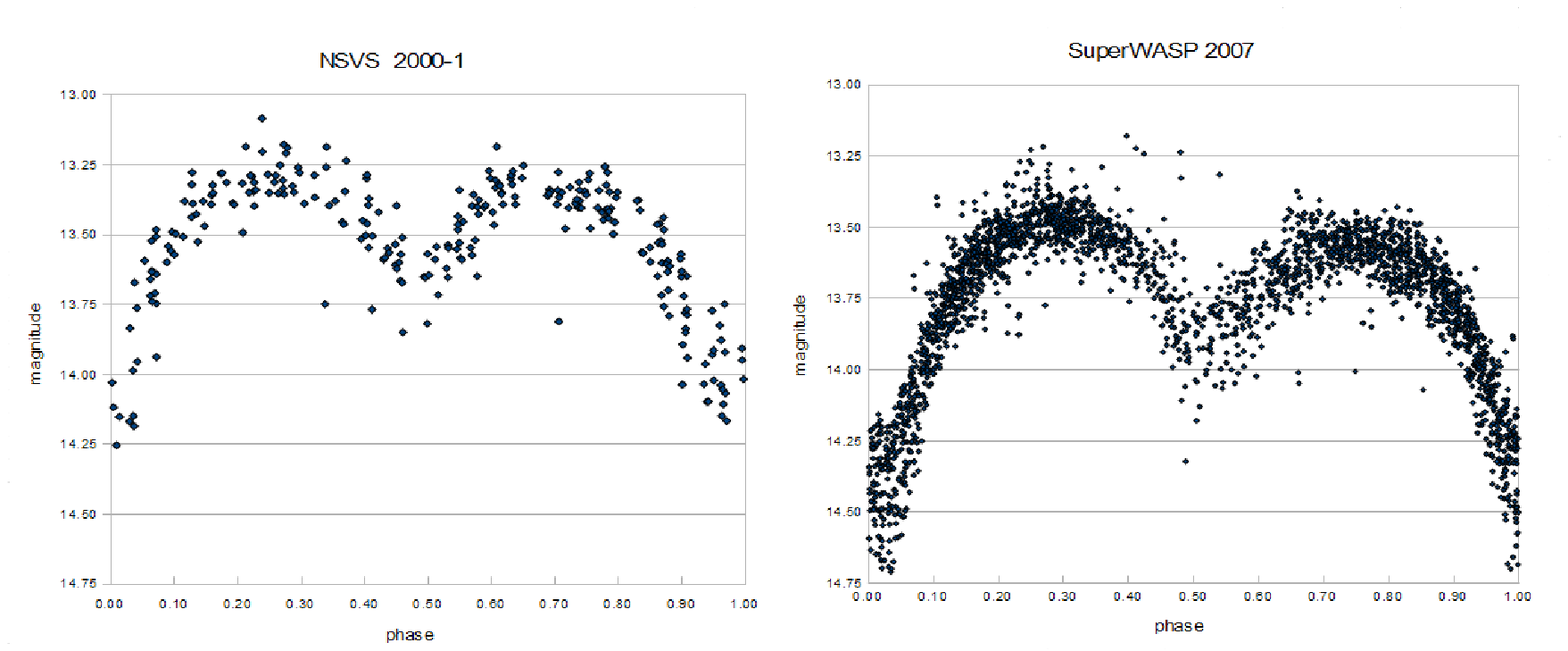}{Light Curves from a) NSVS (left) and b) SuperWASP datasets.}

\vspace*{-0.3cm}

In order to better understand the ephemeris of RGS, we calculated the (O-C) residuals for the times of minimum they derived from their data; see Table 4.  The fourth column gives the residuals for the timings given in RGS; for entries 3-6 we noticed a discrepancy of about two hours, and realized that, since the period is about 11 hours, changing the HJD to one day earlier (and the cycle count two cycles earlier), we could reduce the residuals to just a few minutes.  We speculate that the JD for the UT-date was used as if it were for the calendar date.  Since the calendar dates given in RGS do not correspond to these JD's, we cannot tell just what happened.  Also, even though timings 1 and 2 seem to agree well with our ephemeris, no source data was cited in RGS, so we did not use these timings either.

\begin{table}[!ht]
\small
\centerline{{\bf Table 4.} $O-C$ Linear Residuals for timeings from RGS, based on our Eq. 1}
\begin{center}
\begin{tabular}{cccrrl}
\hline
     No   &      Epoch (Obs) &   Cycle  &  (Obs-Calc) &   (Obs-Calc)* &  Comment\\
          &       HJD        &          &   minutes   &    minutes    &       \\
\hline
    1   &    2454408.9555  & -2499.0  &    7.32     &      ---      &   Primary\\
    2   &    2454427.9001  & -2458.0  &    6.46     &      ---      &   Primary\\
    3   &    2454829.7566  & -1590.5  &  121.19     &     11.98     &   Secondary\\
    4   &    2454829.9867  & -1590.0  &  119.19     &     10.63     &   Primary\\
    5   &    2454849.6213  & -1547.5  &  114.49     &      5.28     &   Secondary\\
    6   &    2454849.8522  & -1547.0  &  114.29     &      5.08     &   Primary\\
\hline
\end{tabular}
\end{center}
* The residuals in col 5 result from increasing the cycle count by two and decreasing the HJD by 1 day.
\end{table}


A few comments about the first eight entries in Table 2 of RGS are in order; these come from the NSVS dataset we used in figure 3a here.  The Epochs given in RGS seem to have ignored the difference of 0.5 days for the MJD, in which those times are given.  The times of minimum correspond to observations which were particularly faint, but not necessarily exact minima.  However, many times of faint magnitude were ignored, and some were included in spite of the fact that the observation was flagged.  

In conclusion, we find a new ephemeris for V963 Per, and caution that times of secondary minimum should not be included in calculating the ephemeris of this star.  We would also caution that the formal errors of times of minimum, like the ones given in our Table 3, may be meaningless as well, since these times can be affected by starspots. We find that the star varies on a short timescale ($\sim$ one month) at the few percent level in brightness, and most of this variation is at a time in the cycle where the spots invoked by RGS are not visible, casting doubt on their interpretation of the spot characteristics.  Given the nature of the star, its period is likely variable, but we cannot say at this point given the unresolved discrepancies in RGS's timings.  Continued observations are important to improve our understanding of this star's ephemeris and variability.

\vspace*{1.0cm}

ACKNOWLEDGMENTS
We thank Dr. Elizabeth M. Green of Steward Observatory for invaluable suggestions in obtaining and reducing the data from the Mont4K camera, and Steward Observatory for allocation of telescope time.  We also thank Dr. Joel Eaton for comments on the manuscript.  This publication makes use of the data from the Northern Sky Variability Survey created jointly by the Los Alamos National Laboratory and University of Michigan.  We have used data from the SuperWASP public archive in this research.  We thank the anonymous referee for careful reading and useful comments.

\references

Anonymous 2006, IBVS 5700

Butters, O. W.; West, R. G.; Anderson, D. R.; Collier Cameron, A.; Clarkson, W. I.; Enoch, B.; Haswell, C. A.; Hellier, C.; Horne, K.; Joshi, Y.; Kane, S. R.; Lister, T. A.; Maxted, P. F. L.; Parley, N.; Pollacco, D.; Smalley, B.; Street, R. A.; Todd, I.; Wheatley, P. J.; \& Wilson, D. M. 2010, \textit{A\&A} \textbf{520}, L10 (SuperWASP)

Odell A.P.; \& Cushing G.E. 2004, IBVS No. 5514
 
Randall, S. K.; Green, E. M.; Van Groote, V.; Fontaine, G.; Charpinet, S.; Lesser, M.; Brassard, P.; Sugimoto, T.; Chayer, P.; Fay, A.; Wroblewski, P.; Daniel, M.; Story, S.; \& Fitzgerald, T. 2007,  \textit{A\&A} \textbf{476}, 1317

Samec, R. G.; Melton, R. A.; Figg, E. R., Labadorf; C. M., Martin, K. P.; Chamberlain, H. A.; Faulkner, D. R.; \& Van Hamme, W. 2010, {\it Astron J}, {\bf 140}, 1150 (RGS)

Wozniak, P. R.; Vestrand, W. T.; Akerlof, C. W.; Balsano, R;, Bloch, J.; Casperson, D.; Fletcher, S.; Gisler, G.; Kehoe, R.; Kinemuchi, K.; Lee, B. C.; Marshall, S.; McGowan, K. E.; McKay, T. A.; Rykoff, E. S.; Smith, D. A.; Szymanski, J.; \& Wren, J. 2004, \textit{Astron J}, \textbf{127}, 2436 (NSVS)

\endreferences

* IRAF is distributed by the National Optical Astronomy Observatories, which are operated by the Association of Universities for Research in Astronomy, Inc., under cooperative agreement with the National Science Foundation.

\end{document}